\documentclass[iop,apj,tighten]{emulateapj}
\usepackage{apjfonts} 
\usepackage{amsmath,amstext}
\usepackage{bm}
\usepackage[breaklinks,colorlinks,citecolor=blue,linkcolor=magenta]{hyperref} 
\usepackage[all]{hypcap} 
\usepackage[usenames,dvipsnames]{color}
\newcommand{\msun}{\ensuremath \mathrm{M}_{\odot}}
\newcommand{\Rsun}{\ensuremath \mathrm{R}_{\odot}}

\DeclareMathAccent{\dot}    {\mathalpha}{operators}{'137} 
\DeclareMathAccent{\ddot}    {\mathalpha}{operators}{'177} 
\shorttitle{Gravitational wave triple systems}
\shortauthors{Meiron, Kocsis, \& Loeb}

\begin{document}
\title{Detecting triple systems with gravitational wave observations}
\author{Yohai Meiron\altaffilmark{1}, Bence Kocsis\altaffilmark{1}, and Abraham Loeb\altaffilmark{2}}
\affil{$^1$Institute of Physics, E\"otv\"os University, P\'azm\'any P. s. 1/A, Budapest, 1117, Hungary;}
\affil{$^2$Center for Astrophysics, 60 Garden Street, MS-51, Cambridge, MA, 02138, USA}

\begin{abstract}
The Laser Interferometer Gravitational Wave Observatory (LIGO) has recently discovered gravitational waves (GWs) emitted by merging black hole binaries. We examine whether future GW detections may identify triple companions of merging binaries. Such a triple companion causes variations in the GW signal due to (1) the varying path length along the line of sight during the orbit around the center of mass, (2) relativistic beaming, Doppler, and gravitational redshift, (3) the variation of the ``light''-travel time in the gravitational field of the triple companion, and (4) secular variations of the orbital elements. We find that the prospects for detecting the triple companion are the highest for low-mass compact object binaries which spend the longest time in the LIGO frequency band. In particular, for merging neutron star binaries, LIGO may detect a white dwarf or M-dwarf perturber at signal to noise ratio of 8, if it is within $0.4\,\Rsun$ distance from the binary and the system is within a distance of 100 Mpc. Stellar mass (supermassive) black hole perturbers may be detected at a factor $5\times$ ($10^3\times$) larger separations. Such pertubers in orbit around the merging binary emit GWs at frequencies above 1 mHz detectable by the Laser Interferometer Space Antenna (\textit{LISA}) in coincidence.
\end{abstract}


\maketitle

\section{Introduction}\label{sec:introduction}
The Laser Interferometer Gravitational Wave Observatory\footnote{\url{http://www.ligo.org/}} (LIGO) 
has recently announced the detection of gravitational waves (GWs) from two merging black hole (BH) binaries GW150914 \citep{2016PhRvL.116f1102A} and LVT151012 \citep{LVTpaper}, opening the era of gravitational wave astronomy. With the further development of GW detectors including VIRGO\footnote{\url{http://www.virgo-gw.eu/}} and KAGRA\footnote{\url{http://gwcenter.icrr.u-tokyo.ac.jp/en/}}, merging compact object binaries are expected to be detected regularly at a rate of several per day within a distance of $7\,$Gpc \citep{LIGOrate}. 
In this work, we examine whether the presence of a triple companion could be detected by measuring the GW signal of the merging binary. 

\begin{figure}
\includegraphics[width=1\columnwidth]{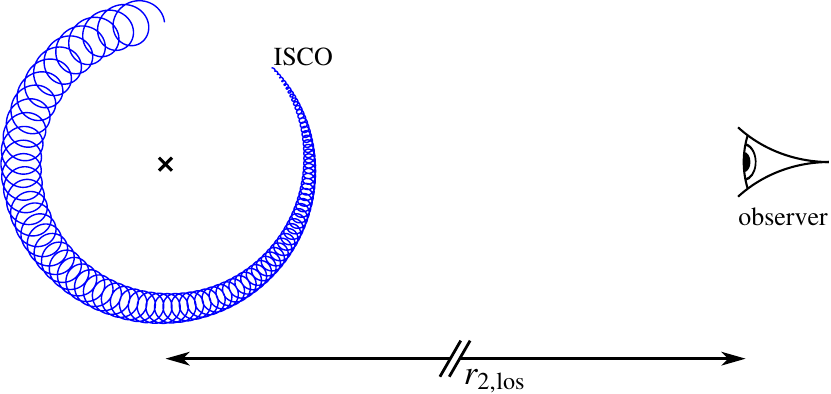}
\caption{\label{fig:schematic} A schematic illustration of the triple systems under consideration (not to scale). The blue curve represents the trajectory of one component of the inspiraling ``inner binary''. The inner binary separation $a_1$ shrinks because of gravitational wave emission, while its center of mass orbits around the triple system's center of mass (marked with a cross) forming an ``outer binary'' with separation $a_2\gg a_1$. The blue curve starts when the inner binary's gravitational wave (GW) frequency enters the detector's sensitive range, and ends when the inner binary reaches the innermost stable circular orbit (ISCO) and coalesces. Depending on the three masses and $a_2$, the inner binary may complete thousands of inner orbits and multiple revolutions around the triple's center of mass while in the LIGO/VIRGO band, whereas for others it completes only a small fraction of an outer orbit (see Equation~\ref{eq:phi2}).}
\end{figure}

We consider a hierarchical triple system, where two stellar mass compact objects form an ``inner binary'', with a triple companion at a large distance compared to the inner binary's orbital separation. The inner binary's orbit shrinks due to GW radiation reaction while it is orbiting the triple system's center of mass (see Figure \ref{fig:schematic}). This orbital motion due to the third companion causes modifications to the GW signal due to a time dependent change in the: (1) path length to the observer, (2) relativistic Doppler and gravitational redshift, and (3) ``light''-travel time of the GW signal as it crosses the gravitational field of the companion. These effects are well studied in pulsar binaries, in which the pulsar orbits another compact object, known respectively as Roemer, Einstein, and Shapiro-delays. Here, the pulsar is replaced by the GW source, the merging inner binary, and instead of timing the radio pulses we measure the distortion of the GW waveform, relative to a theoretical waveform corresponding to an isolated (i.e. unperturbed) inspiraling BH binary. Furthermore, the relativistic beaming of the orbit of the GW source provides an amplitude modulation, and the triple companion has a dynamical influence which drive variations of the intrinsic orbital elements of the GW-emitting binary. In this study, we examine whether any of these effects may be detected in a GW signal to unveil the presence of a third object in the vicinity of the GW source.

Hierarchical triples are common in astrophysics. More than $40\%$ of stellar systems with a white dwarf (WD) and a short period binary form triples, and generally $42\pm5\%$ of massive stars brighter than 10 mag are in triples \citep{1997AstL...23..727T,2006AJ....131.2986P,2014ApJ...782..101P}. Thus, unless the black holes receive a substantial birth kick at their formation\footnote{only small birth kicks are expected in many cases \citep{2016MNRAS.tmp..294A}}, they may also be expected to commonly reside in triple systems. Only one compact object triple is known to date: a close NS+WD binary orbited by another WD, which was found in the Galactic disk \citep{2014Natur.505..520R}.

The likelihood of finding a triple companion may be different among the different environments in which compact object mergers occur: dense dynamical stellar systems such as globular clusters \citep{2000ApJ...528L..17P,2003ApJ...598..419W,OLeary06,Antonini2014ApJ...781...45A,Rodriguez16,2016arXiv160202809O} or galactic nuclei \citep{OKL09,2012PhRvD..85l3005K}, active galactic nuclei \citep{2016arXiv160203831B,2016arXiv160204226S}, galactic field mergers catalyzed by special modes of stellar evolution \citep{Mandel16,Mandel16b,2016arXiv160204531B,Marchant16} or the first stars \citep{2014MNRAS.442.2963K,2016MNRAS.456.1093K,2016arXiv160305655H,2016arXiv160306921I,2016arXiv160404288D}, cores of massive stars \citep{2013PhRvL.111o1101R,2016ApJ...819L..21L,2016arXiv160300511W}, or dark matter halos comprised of primordial black holes \citep{2016arXiv160300464B,2016arXiv160305234C,2016arXiv160308338S}. Close compact object triples may form through common envelope evolution in galaxies \citep{2014ApJ...781L..13T}. 
Alternatively, they may form dynamically in dense stellar environments where many of the compact object mergers detectable by LIGO/VIRGO may originate  \citep{2000ApJ...528L..17P,2005MNRAS.358..572I,2014ApJ...784...71S,Rodriguez16,2016arXiv160202809O}. In fact, the triple companion may be the cause of the compact object merger itself by driving eccentricity oscillations, the so-called Kozai-Lidov effect \citep{2002ApJ...578..775B,2002ApJ...576..894M,2003ApJ...598..419W,2011PhRvL.107r1101K,2012ApJ...757...27A,2013MNRAS.431.2155N,2013PhRvL.111f1106S,2014MNRAS.439.1079A,Antonini2014ApJ...781...45A,Naoz16}.  

The possibility of measuring the influence of a massive perturber for the Laser Interferometer Space Antenna (\textit{LISA}) extreme mass ratio inspirals (EMRIs) was discussed by \citet{2011PhRvD..83d4030Y}. \citet{2010ApJ...724...39S} showed that a corotation resonance may drag test-particles close to the merger of inspiraling intermediate-mass ratio or EMRI black hole binaries, which if tidally disrupted, may lead to an electromagnetic counterpart (see also \citealt{2015PhRvD..91l4016Y} for generalizations to arbitrary mass ratios, \citealt{2011MNRAS.415.3824S} for other mean motion resonances, and \citealt{2016PhRvD..93f4065Z} for the collinear triple configuration for the scalar-tensor theory of gravity). \citet{2011PhRvD..83h4013G} examined post-Newtonian dynamical effects associated with hierarchical triple systems and found that the triple companion affects the octupole GW radiation waveform. Furthermore, \citet{2013ApJ...763..122K} showed that a supermassive black hole in the vicinity of a LIGO/VIRGO source may result in a GW echo detectable with LIGO/VIRGO if the primary signal has a high signal to noise ratio. 
 
More generally, detecting the astrophysical environment of GW sources may be important for understanding their origin.
A shift of the BH ringdown frequency could be used to look for a Planck-density firewall near the horizon with \textit{LISA} and perhaps also with LIGO/VIRGO \citep{2014PhRvD..89j4059B}, if it exists. The influence of an embedding gaseous disk during a GW-inspiral may be detected with \textit{LISA} \citep{2008PhRvD..77j4027B,2011PhRvD..84b4032K,2011PhRvL.107q1103Y} or with the aid of electromagnetic observations 
\citep{2006ApJ...637...27K,2008PhRvL.101d1101K,2012MNRAS.427.2680K,2012ApJ...752L..15G,2012PhRvL.109v1102F,2012ApJ...755...51N,2012ApJ...744...45B,2012ApJ...754...36A,2014PhRvD..90j4030G,2013MNRAS.432.1468M,2015MNRAS.447L..80F}. A double jet may also be characteristic of a binary merger in a gaseous environment \citep{2010Sci...329..927P}. Similar effects may be detected for pulsar timing array (PTA) GW sources \citep{2011MNRAS.411.1467K,2013CQGra..30v4012T,2014PhRvD..89f4060G,2014ApJ...785..115R,2014MNRAS.443L..64G}. The measurement of the interaction of GWs with matter is expected to be practically very challenging \citep{2008ApJ...684..870K,2012MNRAS.425.2407L,2014MNRAS.445L..74M}.

In this work, we quantify the parameter space of triple companion mass and separation where its effect on the GW signal may be detected with Advanced LIGO/VIRGO. 
In Section~\ref{sec:characteristics}, we list the basic equations that define the characteristics of the triple system and the GWs. 
In Section~\ref{sec:detection}, we review the signal-to-noise of detecting the perturbation and the significance of the GW phase shift. 
In Section~\ref{sec:phaseshift}, we analytically estimate the order of magnitude of the GW phase shift for the various physical effects as a function of the triple's physical parameters and present numerical results for the detectability of the third object.
Finally, in Section~\ref{sec:discussion}, we discuss the implications of triple detections using LIGO/VIRGO and \textit{LISA}.

We use geometrized units with ${\rm G}={\rm c}=1$. To change from mass to distance or time units, one should multiply all mass terms by $\rm G/c^2$ or $\rm G/c^3$, respectively.

\section{Characteristics of the triple}\label{sec:characteristics}
We assume a hierarchical triple of mass $(m_\mathrm{a}, m_\mathrm{b}, m_\mathrm{c})$ which consists of an ``inner binary'' and an ``outer binary'' labeled by index 1 and 2, respectively. The inner binary is comprised of $(m_\mathrm{a}, m_\mathrm{b})$, with total mass $M_1=m_\mathrm{a}+m_\mathrm{b}$, symmetric mass ratio $\eta_1=m_\mathrm{a}m_\mathrm{b}/(m_\mathrm{a}+m_\mathrm{b})^2$, and separation $a_1$ which shrinks as a function of time due to GW radiation reaction. The outer binary consists of the center of mass of the inner binary of mass $M_1$ and the outer perturber $m_\mathrm{c}$ with total mass $M_2=m_\mathrm{a}+m_\mathrm{b}+m_\mathrm{c}$, symmetric mass ratio $\eta_2=(m_\mathrm{a}+m_\mathrm{b})m_c/(m_\mathrm{a}+m_\mathrm{b}+m_\mathrm{c})^2$, and separation $a_2$ which may also slowly shrink due to GW radiation reaction. The orbital angular frequency for both binaries $i=(1,2)$  is to leading order
\begin{equation}\label{eq:Kepler}
\Omega_i = M_i^{-1} \left(\frac{a_i}{M_i}\right)^{-3/2}\,.
\end{equation}
For circular orbits, the comoving GW frequency is twice the orbital frequency, $f_i= \Omega_i/\pi=\pi^{-1}(a_{i}/M_i)^{-3/2}M_i^{-1}$. 

As the binaries emit GWs their separations decrease and GW frequencies increase. For circular binaries, The time to merger from given $a_i$ or $f_i$ is
\begin{equation}\label{eq:tmerge}
t_{i,\rm merge} = \frac{5}{256} \frac{a_i^4 }{\eta_i M_i^3}= 5 (8\pi f_i)^{-8/3} \mathcal{M}_i^{-5/3}\,.
\end{equation}
where $\mathcal{M}_i=\eta_i^{3/5}M_i$ is the chirp mass of the $i^{\rm th}$ binary. The inspiral waveform ends\footnote{For maximally spinning BH binaries, $a_{i,\rm ISCO}=1$--$12\,M_i$, depending on the direction of the spin.} at $a_{i,{\rm ISCO}}=6\,M_i$ BH binary or a maximum GW frequency $f_{i,\rm ISCO}=\pi^{-1}6^{-3/2}M_i^{-1}$, which is around $\sim 1.6\,{\rm kHz}$ for NS-NS binaries. 

We label the orbital phase\footnote{More specifically the true anomaly, but we restrict to circular orbits unless mentioned otherwise.} with $\phi_i$ for the two binaries, which satisfy  $\Omega_i = \mathrm{d}\phi_i/\mathrm{d}t$. 
The GW phase $\Phi_i$ satisfies $\mathrm{d}\Phi_i/\mathrm{d}t = f_i$ and so in a comoving frame with the center of mass of the inner binary,
\begin{equation}\label{eq:phiGW}
\Phi_1 = 2\phi_1 = 2\phi_{1,0} + 2 (8\pi f_1 \mathcal{M}_1)^{-5/3} = 2\phi_{1,0} + 2 \left(\frac{|t-t_0|}{5 \mathcal{M}_1}\right)^{5/8}
\end{equation}
where $f_1= \Omega_1/\pi=\mathrm{d}\Phi_1/\mathrm{d}t$ is the comoving GW frequency of the inner binary, $t$ is the comoving time, $\phi_{1,0}$ is the orbital phase of the coalescence. Note that the GW phase accumulates mainly near the minimum observation frequency $f_\mathrm{min}$.

Similar equations hold for the outer binary. If the observation is short relative to $t_{2,\rm merge}$, we may approximate $\Omega_2$ with a constant and
\begin{equation}\label{eq:phi2}
\phi_2 = \phi_{2,0} + \Omega_2 |t-t_0|\,,
\end{equation}
where $\phi_{2,0}$ is the orbital phase at the time of coalescence of the inner binary, $t_0$. 
The orbital phase completed by the outer binary during which the GW frequency of the inner binary is above $f_\mathrm{min}$ is
\begin{align}\label{eq:phi2tot}
\phi_{2,\rm tot} &= \Omega_2 t_{1,\rm merge} 
= 3.5\, \bar{f}_{\rm min}^{-8/3} (4\eta_1)\bar{M_1}^{-5/3}\bar{M_2}^{1/2}\bar{a}_2^{-3/2}\,,
\end{align}
where barred quantities are measured in some specific units: mass parameters are in $\msun=2\times 10^{33}\,\mathrm{g}$, distances such as $a_2$ are in $\Rsun=7\times 10^{10}\,{\rm cm}$, and frequencies such as $f_\mathrm{min}$ are measured in units of 10\,Hz (the minimum detectable GW frequency for Advanced LIGO).

The line of sight (LOS) distance to the center of mass of the inner binary is
\begin{equation}\label{eq:r1LOS}
r_{1,\rm los} = r_{2,\rm los} - \frac{m_\mathrm{c}}{M_2} a_2 \sin\iota_2 \cos\phi_2 \,,
\end{equation}
where we assume that the center of mass of the outer binary is fixed at $r_{2,\rm los}$, and $\iota_2$ is the angle between the orbital angular momentum vector of the outer binary and the line of sight.\footnote{If the center of mass of the merging binary moves at a fixed LOS velocity, $v_1$, then the GW signal changes only by rescaling all mass parameters by a Doppler factor $(1+v_2)$, and rescaling the source distance due to relativistic beaming.} The magnitude of the orbital velocity of the center of mass of the inner binary is
\begin{equation}\label{eq:v1}
v_1 = \frac{m_\mathrm{c}}{M_2} a_2 \Omega_2 =  \frac{m_\mathrm{c}}{M_2} \left(\frac{a_2}{M_2}\right)^{-1/2} \,,
\end{equation}
which is less than $0.03$ (of the speed of light) if $a_2 \geq 10^3 M_2$, the case we are considering here.

We note that the triple must be hierarchical and the inner binary must not be disrupted by the outer binary for these estimates.
The \citet{1995ApJ...455..640E} stability criterion for circular orbits is 
\begin{equation}\label{eq:stability}
\frac{a_2}{a_1}  \gtrsim Y_0 \equiv 1 + \frac{3.7}{q_2^{1/3}} + \frac{2.2}{1+q_2^{1/3}}+  \frac{1.4}{q_1^{1/3}} \frac{q_2^{1/3}-1}{q_2^{1/3}+1}
\end{equation}
where $q_1=m_\mathrm{a}/m_\mathrm{b} \geq 1$ and $q_2=(m_\mathrm{a}+m_\mathrm{b})/m_\mathrm{c}$ are the mass ratio of the inner and outer binary, respectively. 
Furthermore, if the outer object is not a black hole or a neutron star, it needs to be beyond the tidal disruption radius to form a stable triple
\begin{equation}
a_2 \geq r_{\rm tidal} = \left(\frac{3M_1}{4\pi\rho_c} \right)^{1/3} = 
1.2\,\Rsun\, \bar{M}_1^{1/3} \bar{\rho}_c^{\,-1/3}
\end{equation}
where $\bar{\rho}_c$ is the density of the outer object in units of ${\rm g}\,{\rm cm}^{-3}$. Note that for white dwarfs, $\rho \sim 10^6\,{\rm g}\,{\rm cm}^{-3}$, implying that $r_{\rm tidal} \sim 0.01\bar{M}_1^{1/3}\,\Rsun$.

A relativistic triple system which is hierarchical and stable may not have been so in the past. By applying the \citet{1964PhRv..136.1224P} formula for orbital decay to both inner and outer binaries, we find that the ratio of semi-major axis ratio evolves according to\footnote{We assume that the binaries evolve independently from one another and only due to GW emission.}
\begin{equation}\label{eq:at}
\frac{a_2}{a_1} =  \left[ \kappa^4 + \left(\frac{a_{2,0}}{a_1}\right)^4\right]^{1/4}\,,
\end{equation}
where $a_{2,0}$ is the outer binary separation at the merger of the inner binary, $a_1$ is a monotonically decreasing function of time \citep[given by][]{1964PhRv..136.1224P}, and
\begin{equation}
\kappa^4 = \frac{\eta_{2}M_{2}^{3}}{\eta_{1}M_{1}^{3}} = \frac{(m_\mathrm{a}+m_\mathrm{b}+m_\mathrm{c})m_\mathrm{c}}{m_\mathrm{a} m_\mathrm{b}}\,.
\end{equation}
Under the assumption that the inner binary merges first\footnote{Under arbitrary initial conditions, the outer binary may catch up with the inner binary before it merges, disrupting the hierarchical structure and stability.}, the ratio $a_2/a_1$ is monotonically increasing with time, meaning that the triple system becomes more stable as both inner and outer components lose orbital energy to GWs. Thus, one may ask whether there was some point in time that the system has been dynamically unstable, according to some criterion such as Equation (\ref{eq:stability}). Depending on $m_\mathrm{c}$, $\kappa$ may be arbitrarily small or large. In case $\kappa\geq Y_0$, the triple remains dynamically stable forever in the past of the inner binary merger, approaching an asymptotic self-similar stationary state\footnote{In the limit $a_{2,0}\approx 0$, the triple evolves self-similarly down all the way to ISCO of either binaries.} with $a_2/a_1\approx \kappa$. Otherwise if $\kappa<Y_0$, then the triple becomes dynamically unstable in a time 
\begin{equation}
t_{\rm stable} = 6.0\times 10^8\,{\rm yr}\,(4\eta_1)^{-1} \bar{M}_1^{-3}  (Y_0^4 - \kappa^4)^{-1} (\bar{a}_{2,0})^4 \quad{\rm if}~\kappa < Y_0\,
\end{equation}
before the merger of the inner binary.
The characteristic past lifetime or residence time of a circular triple with $a_2\sim a_{2,0}$ is the minimum of $t_{\rm stable}$ and $t_{2,\rm merge}$ (Equation~\ref{eq:tmerge}). The likelihood of finding a triple companion at $a_{2,0}$ is proportional to this characteristic timescale. We note that these estimates are significantly modified for eccentric triples.

\section{Detecting GW Perturbations}\label{sec:detection}

\begin{figure*}
\includegraphics[width=0.5\textwidth]{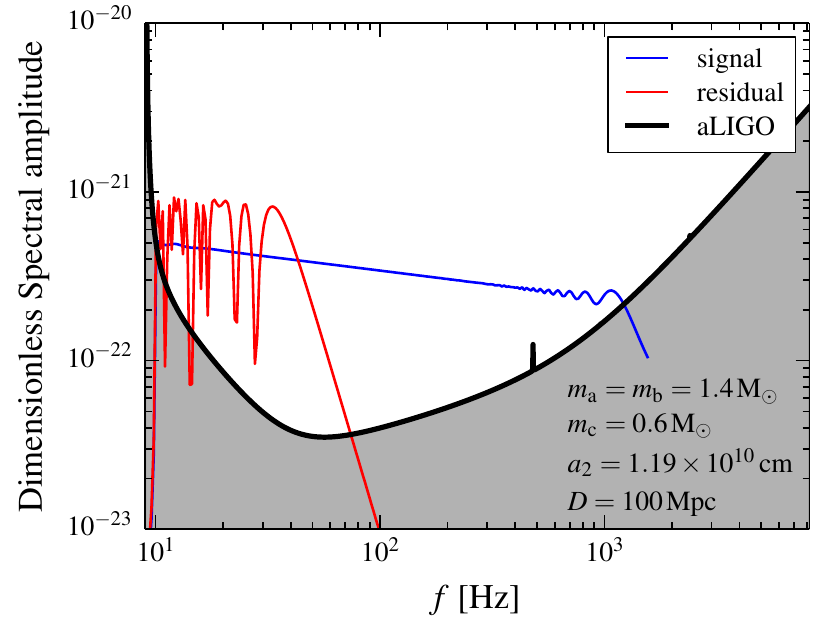}
\includegraphics[width=0.5\textwidth]{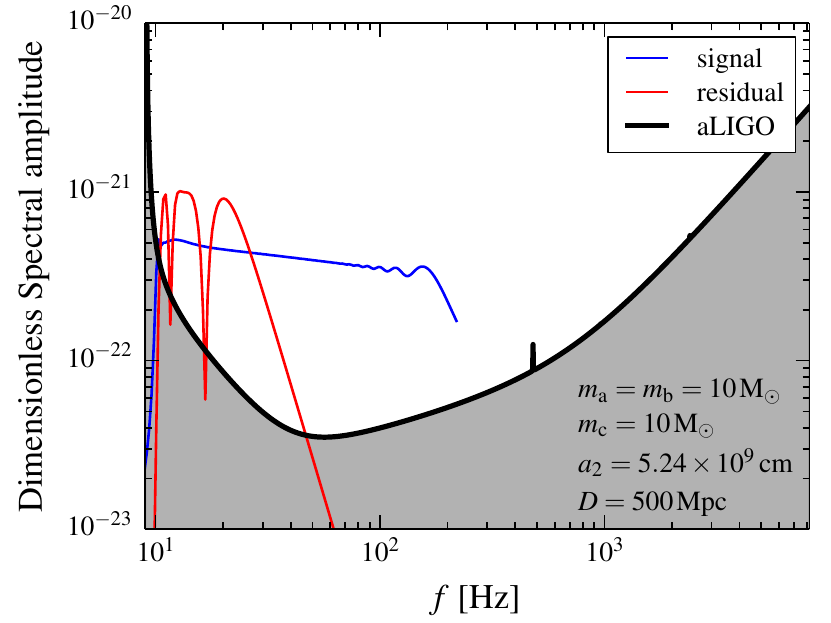}
\caption{\label{fig:spectrum} The dimensionless gravitational wave spectral amplitude (blue curve) and root mean square (rms) spectral noise amplitude per logarithmic frequency bin of Advanced LIGO (black curve) from a merging binary system of (\textit{left panel}) two $1.4\,\msun$ neutron stars (NSs) with a $0.6\,\msun$ white dwarf companion on a circular orbit at a separation $1.185\times 10^{10}$\,cm located at a distance of 100 Mpc from the Earth; (\textit{right panel}) two $10\,\msun$ black holes (BHs) with another $10\,\msun$ BH companion on a circular orbit at a separation $5.244\times 10^9$\,cm located at a distance of 500 Mpc from the Earth. The red curves show the spectral density of the residual between the signal and a reference signal of a merging binary with no triple companion. The signal progresses in time from left to right; the line ends when the (inner) binary reaches the innermost stable circular orbit (ISCO). The coalescence and ringdown phases are not shown.}
\end{figure*}

For an inspiraling source at a fixed distance\footnote{If the source is at a cosmological redshift $z$, $D$ is the luminosity distance and the mass parameters must be multiplied by $(1+z)$. } $D$ and a random\footnote{The prefactor assumes a root-mean-square average of the detected GW strain in a single LIGO-type detector for isotropically chosen source sky position and orientation. We neglect the effects of a peculiar velocity and weak lensing here \citep{2006ApJ...637...27K}. } source sky position and orientation, the detected dimensionless strain is 
\begin{equation}
h(t) = \frac{16}{5} \frac{\eta_1 M_1}{D} f(t)^{2/3} \cos [\Phi(t)]
\end{equation}
where $\Phi(t)$ is given by Equation~(\ref{eq:phiGW}) and $f=\mathrm{d}\Phi/\mathrm{d}t$, and the one-sided Fourier transform in the stationary phase approximation is to leading (2.5 post-Newtonian) order \citep{1994PhRvD..49.2658C}
\begin{equation}
\tilde h = \frac{\mathcal{M}_1^{5/6}}{\pi^{2/3} \sqrt{30}D} f^{-7/6} e^{i\Psi(f)}
\end{equation}
and
\begin{align}\label{eq:Psi}
\Psi(f) &= 2\pi f t(f) -\Phi(f) - \frac{\pi}{4} \nonumber\\
&= 2\pi f t_0 - 2\phi_{1,0} + \frac{3}{4}(8\pi \mathcal{M}_1 f)^{-5/3}- \frac{\pi}{4} \,.
\end{align}
In the second line we have used Equation~(\ref{eq:phiGW}) for the inner binary, where $\phi_{1,0}$ is the orbital phase at merger, and $t(f) = t_0 + t_{1,\rm merge}(f)$ given by Equation~(\ref{eq:tmerge}).  

We discuss the detectability of a GW perturbation following \citet{2011PhRvD..84b4032K}. 
To detect a perturbation to the GW signal, $\delta h$, the signal-to-noise ratio (SNR) of the  perturbation 
\begin{align}
\left\langle\frac{S_{\rm pert}^2}{N^2}\right\rangle &= 
4\int_{f_\mathrm{min}}^{f_{\max}}\frac{|\delta\tilde{h}|^2}{S_n} \mathrm{d}f =
8\int_{f_\mathrm{min}}^{f_{\max}}\frac{|\tilde{h}|^2 (1-\cos \delta \Psi)}{S_n} \mathrm{d}f 
\nonumber\\& \approx 4\int_{f_\mathrm{min}}^{f_{\max}}\frac{|\tilde{h}|^2}{S_n}  \delta\Psi^2 \mathrm{d}f
\label{eq:SNR}
\end{align}
must exceed a given detection threshold, typically $S/N\gtrsim 8$ for a false alarm probability of 0.02 \citep{LVTpaper}. Here $S_n$ is the one-sided mean-square spectral noise density with units of $1/{\rm Hz}$ characteristic of the instrument \citep{LIGO_sensitivity}, and for Advanced LIGO $f_\mathrm{min}\sim 10$\,Hz, $f_{\max}$ is the maximum frequency set by the coalescence, and $\delta \Psi$ is the dephasing caused by the perturbation. 
In the second equality in Equation~(\ref{eq:SNR}), we assumed that the GW signal $h$ is perturbed by a GW phase, $\delta \Phi$, which leads to a corresponding Fourier phase shift $\delta \Psi$ (Equation~\ref{eq:Psi}), so $\delta \tilde{h} = \tilde{h} e^{i \delta \Psi} - \tilde{h}$, and in the third equality we expanded to second order in $\delta \Psi$.

The conclusion from Eqs.~(\ref{eq:Psi}) and (\ref{eq:SNR}) is that the perturbation may be detected if the original unperturbed GW source has $S/N\gtrsim 8$ and the perturbation generates a phase shift $\delta \Phi \gtrsim 1\,$rad.\footnote{$\delta \Phi$ may be somewhat smaller if the unperturbed signal has $S/N\gg 8$.} 
Note that the phase shift is an intrinsic property of the perturbation, independent of the source distance from the Earth.

Figure~\ref{fig:spectrum} shows the dimensionless spectral amplitude of the signal $2f\tilde{h}$ and the residual $2f\delta\tilde{h}$ in blue and red lines respectively, and the root-mean-square noise per logarithmic frequency bin ($\sqrt{f S_n}$, black curve, \citealt{LIGO_sensitivity}). Specifically, the red curve shows the perturbation of the signal due to the leading order effect of a third companion, the Doppler phase, discussed in Section~\ref{sec:doppler} below. The cases shown are (left panel) two neutron stars inspiraling in the presence of a white dwarf, and (right panel) two BHs inspiraling in the presence of a third BH of the same mass (right panel). The ratio of the blue and the black curves integrated over $\ln f$ gives the $S/N$.

\section{Perturbations of GWs in triple systems}\label{sec:phaseshift}

In the following subsections, we calculate the GW phase shift corresponding to the various physical effects related to the triple companion. 

\subsection{Doppler shift}\label{sec:doppler}

\begin{figure*}
\includegraphics[width=0.5\textwidth]{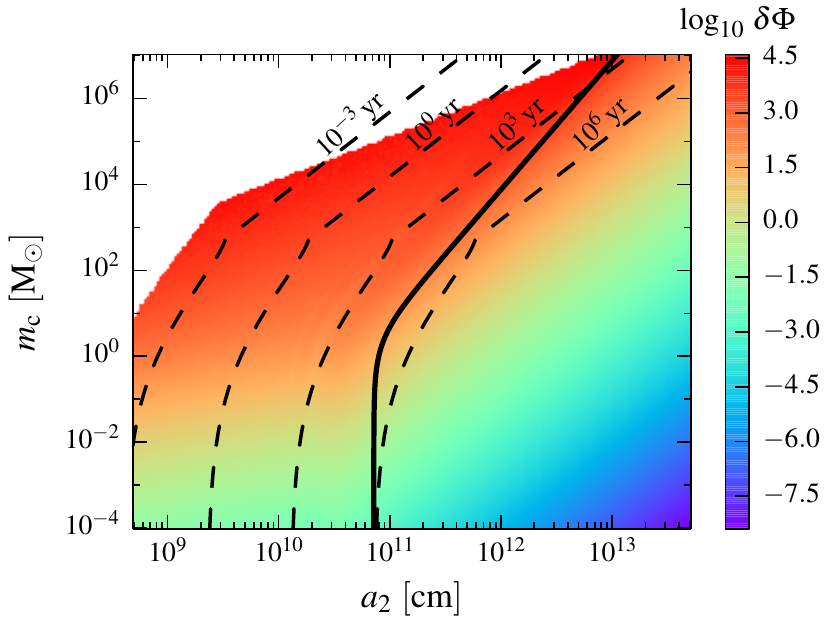}
\includegraphics[width=0.5\textwidth]{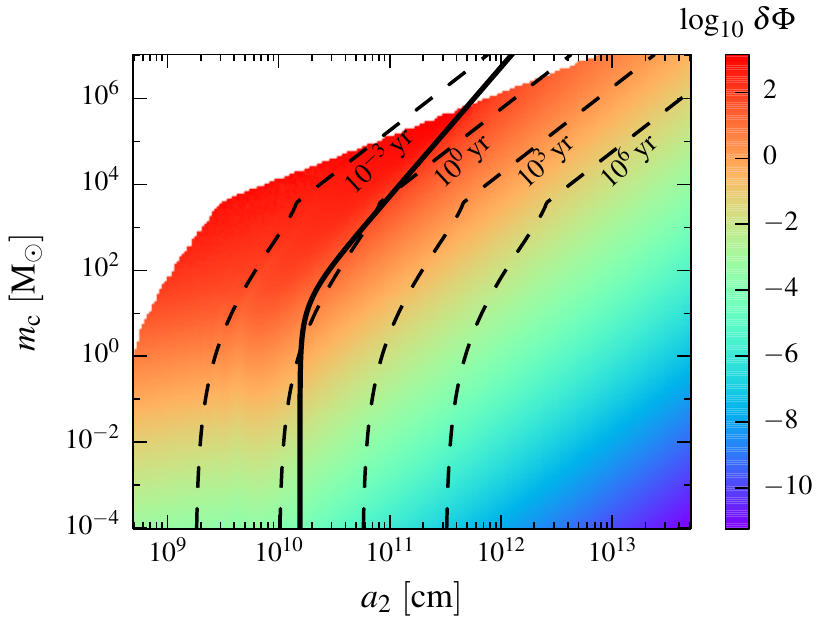}
\caption{\label{fig:dopplerphase} The total Doppler phase shift of the GW signal compared to a reference waveform, from a merging NS-NS binary (\textit{left panel}) and a BH-BH binary (\textit{right panel}) due to a perturber, while the signal is in the LIGO/VIRGO frequency band; $\log_{10}\delta \Phi$ is shown as a function of the perturber's distance and mass. For systems along the thick solid line, the outer binary has completed one radian of its orbit while the system is in the LIGO band (left of the curve means larger fraction of the orbit). The triple system is unstable in the white region on the top left due to either Newtonian dynamical reasons or the outer separation is smaller than the ISCO; the dashed lines represent the time that the system could have been dynamically stable (i.e. has existed no longer than the amount of time shown on the line). The signal may be detectable if $\log_{10}\delta \Phi \gtrsim 0$ and the source is within the LIGO horizon. The only LIGO-specific information that enters into this figure is $f_\mathrm{min}=10\,\mathrm{Hz}$ as it determined the signal's duration, the phase shift is otherwise independent of detector properties and the source's distance.}
\end{figure*}
\begin{figure*}
\includegraphics[width=0.5\textwidth]{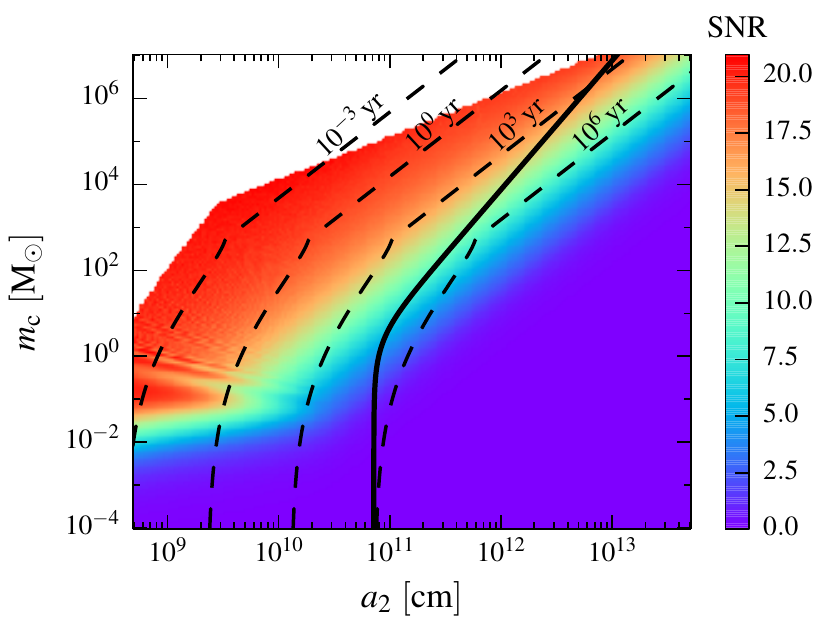}
\includegraphics[width=0.5\textwidth]{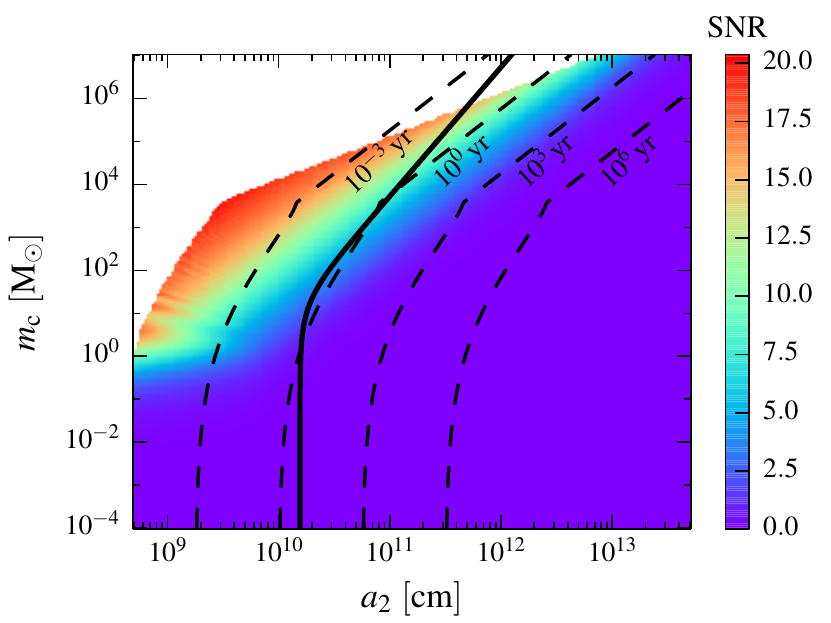}
\caption{\label{fig:SNR} Similar to Figure \ref{fig:dopplerphase} but now showing the signal-to-noise ratio of the residual between the signal and the reference waveform. The Advanced LIGO sensitivity curve is used, and the NS-NS binary (\textit{left panel}) is put at $100\,\rm Mpc$ from the Earth, while the BH-BH binary (\textit{right panel}) is put at a distance of $500\,\rm Mpc$.
}
\end{figure*}

In analogy with signals emitted by pulsars in binary systems, we expect the largest perturbation in the GW signal to be caused by the variation of the line-of-sight distance to the source, which leads to a shift in the arrival time of pulses called Roemer (or R{\o}mer) delay. We are interested in comparing the GW signal of the actual inspiraling inner binary orbiting in a hierarchical triple system (the \textit{source}), and a fictitious isolated inspiraling binary (the \textit{reference system}). This reference system has a constant center of mass (line-of-sight) velocity $v_{1,\rm ref}$ and merges at the same time as the source, where the position, velocity and phase are set to equal those of the source. The GW phase difference is
\begin{equation}\label{eq:phiD}
\delta \Phi_{\rm D} = \Phi[t_{\rm em}-r_{1,\rm los}(t_{\rm em})] - \Phi[(1-v_{1,\rm ref})t_{\rm em}-r_{1,\rm los}(t_0)]\,,
\end{equation}
where $\Phi(t)$ is given by Equation~(\ref{eq:phiGW}), $t_{\rm em}$ is time of emission (retarded time),  $r_{1,\rm los}$ is the line-of-sight distance given by Equation~(\ref{eq:r1LOS}), and $t_0$ is the time of merger. In the following calculations we set $v_{1,\rm ref}$ to be the value of the line of sight velocity at merger, $v_{1,\rm ref}=\dot{r}_{1,\rm los}(t_0)$. 

Before showing the numerical result for Equation~(\ref{eq:phiD}) for various triple-system parameters, it is useful to get a rough estimate of the order of magnitude of this effect analytically.
\citet{2011PhRvD..83d4030Y} examined the GW phase shift when the outer binary orbital phase is small enough that the term $r_{1,\rm los}(t_{\rm em})$ may be approximated by its quadratic Taylor series at $t_0$
\begin{equation}\label{eq:r1LOS2}
r_{1,\rm los}(t_{\rm em}) \approx r_{1,\rm los}(t_0) +v_{1,\rm los} t_{\rm em}  + \frac12 \dot{v}_{1,\rm los} t_{\rm em}^2 \,,
\end{equation}
where the velocity and acceleration may be calculated from Equation~(\ref{eq:r1LOS}) to be
$v_{1,\rm los} = - m_\mathrm{c} M_2^{-1} a_2 \Omega_2 \sin\iota_2 \sin\phi_{2}$ and 
$\dot{v}_{1,\rm los} = - m_\mathrm{c} M_2^{-1} a_2 \Omega_2^2 \sin\iota_2 \cos\phi_{2}$.
Setting $v_{1,\rm ref}=v_{1,\rm los}(t_0)$, the phase difference $\delta \Phi_{\rm D}$ is zero at time $t_0$ due to the definition of $\Phi(t)$ in Equation~(\ref{eq:phiGW}), and it accumulates during the GW observation for earlier times before merger. After expanding both $\Phi(t)$ terms in Equation~(\ref{eq:phiD}) in a series to first order in its argument around the point $t=r_{1,\rm los}(t_0) +v_{1,\rm los} t_{\rm em}$ using Equation~(\ref{eq:r1LOS2}), the first two terms drop out in Equation~(\ref{eq:r1LOS2}) and only the term proportional to $\dot{v}_{1,\rm los}$ remains in the phase difference. We get that the total Doppler phase shift during the full LIGO measurement from GW frequency $f_{\min}$ to merger is 
\begin{align}\label{eq:phiDsimple}
\delta \Phi_{\rm D} &\approx \frac{1}{2} \dot\Phi \dot{v}_1 t_{1,\rm merge}^2
=
\frac{1}{2}f_\mathrm{min} \dot{v}_1 t_{1,\rm merge}^2 \nonumber\\&=
890\, \bar{f}_{\rm min}^{-13/3} (4\eta_1)^{-2}\bar{M}_1^{-10/3} \bar{m}_\mathrm{c} \bar{a}_2^{-2}   \sin\iota_2 \cos\phi_{2}
\,,
\end{align}
where we used Equation~(\ref{eq:tmerge}) for $t_{1,\rm merge}$. The quantities denoted by a bar are in physical units and are defined under Equation~(\ref{eq:phi2tot}).
This approximation assumes that $\phi_2$ is approximately constant during the measurement. Let us examine where this holds, the distance $a_2$ where $\phi_2$ changes by less than 1 radian while the inner binary is in the LIGO band, from Equation~(\ref{eq:phi2tot}) is,
\begin{align}\label{eq:a2onerad}
a_{2} \gtrsim & 
\left\{
\begin{array}{ll}
2.3\,\Rsun\, \bar{f}_\mathrm{min}^{-16/9}\, (4\eta_1)^{-2/3} \bar{M}_1^{-7/9}~&{\rm if}~ m_\mathrm{c}\ll M_1\,,\\
2.3\,\Rsun\, \bar{f}_\mathrm{min}^{-16/9}\, (4\eta_1)^{-2/3} \bar{M}_1^{-10/9} \bar{m}_\mathrm{c}^{1/3}~&{\rm if}~ m_\mathrm{c}\gg M_1\,.
\end{array}
\right.
\end{align}
If this condition is not satisfied, then the outer binary makes a larger revolution during the LIGO measurement than 1 radian, and the simple estimate in Equation~(\ref{eq:phiDsimple}) becomes inaccurate. For 1 radian outer binary revolution,
\begin{align}\label{eq:phioneradian}
\delta\Phi_{\rm D}^{(1\,\mathrm{rad})} \approx
\left\{
\begin{array}{ll}
170\sin\iota_2\, \bar{f}_\mathrm{min}^{-7/9}\, (4\eta_1)^{-2/3} \bar{M}_1^{-16/9} \bar{m}_\mathrm{c}~&{\rm if}~ m_\mathrm{c}\ll M_1\,,\\
170\sin\iota_2\, \bar{f}_\mathrm{min}^{-7/9}\, (4\eta_1)^{-2/3} \bar{M}_1^{-10/9} \bar{m}_\mathrm{c}^{1/3}~&{\rm if}~ m_\mathrm{c}\gg M_1\,,
\end{array}
\right.
\end{align}
Furthermore, note that $a_2$ must be greater than the bound set by the hierarchical triple stability criterion in Equation~(\ref{eq:stability}) where $a_1 = 5.1\times 10^7 {\,\rm cm}\, \bar{M}_1^{1/3} \bar{f}_\mathrm{min}^{-2/3}$ when the inner binary enters the LIGO band (see Equation~\ref{eq:Kepler}), and the inner binary must be outside of the ISCO of the outer binary, $a_1\geq a_{2,\rm ISCO}$. 
Equations~(\ref{eq:phiDsimple}) and (\ref{eq:phioneradian}) show that GW phase shift due to the third object may be significant for a wide range of parameters. 

Figure~\ref{fig:dopplerphase} shows the Doppler phase shift for merging NS-NS and BH-BH binaries as a function of the triple companion distance and mass, by fully numerically solving Equation~(\ref{eq:phiD}). The black dashed lines show the amount of time the system has been hierarchical and stable as discussed in Section~\ref{sec:characteristics}, $10^{-3,0,3,6}\,$yr from left to right, respectively. The black solid line corresponds to 1 rad orbit for the outer binary during the lifetime in the LIGO band, for reference (cf. Equation~\ref{eq:phioneradian}). The region to the left of the black solid line corresponds to systems where the outer binary completes a larger orbital phase during the LIGO measurement. The only LIGO-specific information that enters into this figure is $f_\mathrm{min}=10\,\mathrm{Hz}$, which determined the signal's duration, the phase shift is otherwise independent of detector properties and the source's distance.

Figure~\ref{fig:SNR} shows the SNR of the residual signal for merging NS-NS and BH-BH binaries as a function of the triple companion distance and mass, using the Advanced LIGO sensitivity curve \citep{LIGO_sensitivity}. The residual signal is the difference between the aforementioned source and reference system. For each pixel in the two panels in the Figure, the full time domain waveform is calculated for both source and reference system, Fourier transformed, and integrated according to Equation~(\ref{eq:SNR}), i.e. without utilizing the stationary phase or the small angle approximations.

\subsection{Gravitational redshift}\label{sec:redshift}
Another potentially important physical effect that distorts the signal in a triple system is gravitational redshift. In the presence of an additional mass such as that of the triple companion's, the gravitational waves from the inner binary have to climb out of a deeper potential well than in the isolated binary case, and are thus redshifted with respect to the observer, in analogy with electromagnetic radiation in the same situation. This is the result of gravitational time dilation or difference in clock rate, which has an equivalent effect on the GW phase as the difference in ``light''-travel time we discussed in Section~\ref{sec:doppler}.

Since a GW-generating binary of mass $M$ at fixed redshift $z$ may not be distinguishable from an isolated binary with mass $M(1+z)$, the only way gravitational redshift can affect the signal in a measurable way is if the outer binary is eccentric, and the amount of redshift changes along the orbit. Thus, for the reference system we do not choose an isolated binary as in the previous Section, but binary at an arbitrary point along the outer eccentric orbit where the redshift is $z_0$. The phase difference is thus
\begin{equation}\label{eq:phiz}
\delta \Phi_{\rm z} = \Phi[t_{\rm em}(1+z(t_{\rm em})] - \Phi[t_{\rm em}(1+z_0)]
\end{equation}
where $\Phi(t)$ is given by Equation~(\ref{eq:phiGW}), $z(t_{\rm em})$ is the gravitational redshift of the GW source corresponding to the distance from $m_\mathrm{c}$ at source time $t_{\rm em}$. To leading order in the small quantity $m_\mathrm{c}/a_2$ the gravitational redshift is
\begin{equation}
z \approx \frac{m_\mathrm{c}}{a_2} \frac{1 + e_2\cos\phi_2}{1-e_2^2} = \frac{m_\mathrm{c}}{a_2(1 - e_2\cos E_2)}
\end{equation}
where $\phi_2$ is the true anomaly and $E_2$ is the eccentric anomaly which evolves as $\Omega_2 t = E_2-\sin E_2$, and we made some simplifying assumptions about the geometry of the system (namely that the eccentricity vector, the angular momentum vector and the line of sight are in the same plane).\footnote{If the outer orbits completes more than one revolution in the LIGO band, GR precession may not be neglected.}

If we set $z_0=z(t_0)$ and substitute Equation~(\ref{eq:phiGW}) in Equation~(\ref{eq:phiz}), the phase difference between the model including redshift and one that assumes a constant redshift, vanishes at merger by definition. The total phase difference accumlates as a function of time before merger. For the full GW observation, Equation~(\ref{eq:phiz}) must be evaluated at the point where the signal enters the sensitive frequency band at $f_{\min}$. Since $z$ is much less than unity, we may expand Equation~(\ref{eq:phiz}) to first order around this point to get
\begin{align}\label{eq:dphiz2}
\delta \phi_{\rm z} &\approx  \dot{\Phi} \,t_{1,\rm merge} \Delta z =  f_{\min} t_{1,\rm merge} \frac{m_\mathrm{c}}{a_2}\frac{e_2 }{1-e_2^2} \Delta \cos\phi_2
\nonumber\\&= 0.74\, \bar{f}_\mathrm{min}^{-5/3} (4\eta_1)^{-1} \bar{M}_1^{-5/3} \bar{m}_\mathrm{c} \bar{a}_2^{-1} \frac{e_2 }{1-e_2^2}\Delta \cos\phi_2
\end{align}
where $\Delta \cos\phi_2$ is the change in $\cos\phi_2$ during the time the source is in the LIGO frequency band (Equation~\ref{eq:tmerge}), which is at maximum 2 if it completes half an orbit. If it completes less than one radian, we can approximate 
$|\Delta \cos\phi_2|\approx \Omega_2 t_{1,\rm merge}\sin\phi_{2,0}$ to leading order around $\phi_{2,0}$, which gives
\begin{equation}
\delta \Phi_{\rm z} \approx
\left\{
\begin{array}{ll}
2.6\chi \bar{f}_\mathrm{\min}^{-13/3}\, (4\eta_1)^{-2} \bar{M}_1^{-17/6} \bar{m}_\mathrm{c} \bar{a}_2^{-5/2}~&{\rm if}~ {m_\mathrm{c}}\ll {M_1}\,,\\
2.6\chi \bar{f}_\mathrm{\min}^{-13/3}\, (4\eta_1)^{-2} \bar{M}_1^{-10/3} \bar{m}_\mathrm{c}^{3/2} \bar{a}_2^{-5/2}~&{\rm if}~ {m_\mathrm{c}}\gg {M_1}\,,
\end{array}
\right.
\end{equation}
where $\chi = [e_2/(1-e_2^2)] \sin \phi_{2,0}$. The maximum phase shift corresponds to the case where the outer binary completes exactly one half orbit, which implies that
\begin{equation}
\delta \Phi_{\rm z} \leq
1.4\frac{e_2}{1-e_2^2}  \bar{f}_\mathrm{min}^{1/9} (4\eta_1)^{-1/3}
\times \left\{
\begin{array}{ll}
 \bar{M}_1^{-8/9} \bar{m}_\mathrm{c} &{\rm if}~ {m_\mathrm{c}}\ll {M_1}
 \,,\\
 \bar{M}_1^{-5/9} \bar{m}_\mathrm{c}^{2/3} &{\rm if}~ {m_\mathrm{c}}\gg {M_1}
 \,.
\end{array}
\right.
\end{equation}
This shows that the variation of the gravitational redshift around the perturber is typically smaller than the Doppler phase, but it may still be several radians if the perturber is a BH with $m_\mathrm{c}\gtrsim 5\msun$. Note that $\delta \Phi_{\rm z}$ is independent of the binary inclination, and it is nonzero only if the perturber is on an eccentric orbit. 

\subsection{Shapiro delay}
The Shapiro delay is a well known general-relativistic effect that causes the delay in arrival time of a signal when it passes in the gravitational field of a massive object. The time shift caused by the signal propagating in the gravitational field of the perturber is given by equation (5.5) in  \citet{1986ARA&A..24..537B},  
\begin{equation}
\delta t_{\rm S} = m_\mathrm{c} \ln \left|\frac{1+e_2\cos\phi_2}{1 - \sin\iota_2\cos (\phi_2+\omega_{2})}\right|
\end{equation}
where $\omega_{2}$ is the outer binary's argument of periastron. The corresponding phase shift may be derived similarly to that presented in Section~\ref{sec:redshift}, which gives to leading order
\begin{align}
\delta \Phi_{\rm S} &= \dot{\Phi} \delta t_{\rm S} = 2\pi f_\mathrm{min} \delta t_{\rm S}
\nonumber\\ &\leq   2\pi f_\mathrm{min} m_\mathrm{c} |\ln\Lambda|  = 3.1\times 10^{-4} \bar{f}_\mathrm{min} \bar{m}_\mathrm{c}|\ln \Lambda|
\end{align}
where $\dot{\Phi}$ is the time-derivative of the GW phase given by Equation~(\ref{eq:phiGW}) evaluated at $t=t_{1,\rm merge}$ given by Equation~(\ref{eq:tmerge}), and in the last line we estimated the maximum value of the Shapiro delay assuming a half orbit of the outer binary where
\begin{equation}
\ln \Lambda = \ln\left(\frac{1+e_2}{1-e_2}\right) + \ln\left(\frac{1-\sin\iota_2}{1+\sin\iota_2}\right)\,.
\end{equation}
The expectation value for thermally distributed eccentricities\footnote{Note that $e_2$ may not approach unity since that would lead to the disruption of the inner binary.} and isotropically distributed inclinations is $\langle |\ln \Lambda| \rangle = 2$. This expression shows that $\delta \Phi_{\rm S}$ is typically much less than 1 radian unless $m_\mathrm{c}\gtrsim 10^3\,\msun$ or if the outer binary is almost exactly edge on. The Doppler shift and the gravitational redshift typically cause larger perturbations.

\subsection{Dynamical effects}\label{sec:dynamical}
In the above Sections, the waveform emitted by the binary was intrinsically unchanged by the presence of the perturber; the gravitational wave signal observed on the Earth was distorted due to the change of frame of reference. Now we examine the dynamical torque generated by the triple companion which may change the orbital elements of the inner binary. The leading-order dynamical perturbation is the quadrupole component of the tidal gravitational field of the perturber \citep{2014PhRvD..89d4043W}. This leads to both oscillatory variations on the inner orbit timescale and a secular change in the eccentricity and angular momentum vector on much longer timescales, discussed next \citep[see][for further post-Newtonian dynamical three body effects for eccentric triples]{2011PhRvD..83h4013G,2013ApJ...773..187N}.

\subsubsection{Nodal precession}
If the inner and outer binaries are not in the same plane, the angular momentum and eccentricity vectors of the inner binary undergo long-duration changes. For a circular inner binary, the angular momentum of the inner binary $\mathbf{L}_1$, precesses around the total angular momentum $\mathbf{L}_{\rm tot}=\mathbf{L}_1+\mathbf{L}_2$. The corresponding nodal precession rate to leading Newtonian quadrupole order is \citep{2013MNRAS.431.2155N,2013ApJ...773..187N}
\begin{equation}
\Omega_{1,\rm nodal} = \frac{3}{4}\frac{m_\mathrm{c}}{M_1}\Omega_1 \frac{a_1^{3}}{a_2^{3}} \frac{L_{\rm tot}}{L_2}\frac{\cos \theta}{(1-e_2^2)^{3/2}}
\end{equation}
where $\Omega_1 = \pi f$, $a_1=M_1 (\pi M_1 f)^{-2/3}$ is the angular frequency and $\mathbf{L}_i=\eta_i M_i^{3/2} (1-e_i^2)^{1/2} \hat{\mathbf{L}}_i$ for binary $i$,  $\hat{\mathbf{L}}_i$ is a unit vector, $L_i=\|\mathbf{L}_i\|$, and $\cos \theta = \hat{\mathbf{L}}_1\cdot \hat{\mathbf{L}}_2$. The orbital plane precession angle is set by
\begin{align}
\delta\varphi &=\frac{L_2}{L_{\rm tot}} \Omega_{1,\rm nodal} t_{1,\rm merge} 
\nonumber\\
&= 5.2\times 10^{-5}\bar{f}^{-11/3} \bar{M_1}^{-5/3} m_c \bar{a}_2^{\,-3} \frac{\cos \theta}{(1-e_2^2)^{3/2}}
\end{align}
Secular precession effects are therefore expected to be significant if $a_2\lesssim 0.1 \Rsun$ for stellar mass perturbers or if $a_2\lesssim 1 \Rsun$ for intermediate or supermassive BH perturbers.

\subsubsection{Change in the orbital shape}
The tidal force of the perturber acting on the binary due to the triple companion affects the orbital shape similar to how the Moon raises ocean tides on Earth. In a corotating frame with angular velocity $\Omega_1$ with the inner binary, the Newtonian equations of motion become
\begin{equation}
\ddot{\mathbf{r}}_1 = \Omega_1^2 \mathbf{r}_1 - 2\,\bm{\Omega}_1\times\dot{\mathbf{r}}_1 - \frac{M_1}{r_1^3}\mathbf{r}_1 - \frac{m_\mathrm{c}}{r_2^3}\mathbf{r}_1 + 3\frac{m_\mathrm{c}(\mathbf{r}_1\cdot \mathbf{r}_2)}{r_2^5}\mathbf{r}_2
\end{equation}
where $\mathbf{r}_i$ is the separation vector of the $i^{\rm th}$ binary, the first two terms are the centrifugal and Coriolis forces for a coplanar triple, and the last two are the tidal force. The mean orbital frequency is modified by the fourth term, and the last term introduces a time-dependent perturbation to the orbital shape. If the unperturbed orbit is approximately circular, the acceleration in the $\mathbf{r}_2$ direction due to the last term is on average $\frac32 m_c r_1/r_2^3$. Assuming a constant acceleration of this magnitude in this direction for a half-period duration, $\pi/\Omega_1$, we may estimate the corresponding distance traveled, and the corresponding orbital eccentricity:
\begin{align}\label{eq:tides}
e_1&\sim \frac{\Delta r_1}{r_1} \sim \frac{1}{2}\times \frac{3}{2} \frac{m_\mathrm{c}}{r_2^3} \times \left(\frac{\pi}{\Omega_1}\right)^2= 10^{-10} \bar{f}^{-2} \bar{m}_\mathrm{c} \bar{r}_2^{\,-3}\,.
\end{align}
The orbital eccentricity also changes the shape of the GW waveform. If a corresponding phase shift is of order $\delta \Phi \sim e_1 \Phi$, this may be significant for LIGO/VIRGO if $r_2\lesssim 0.1 \Rsun$ for stellar mass perturbers or $r_2\lesssim \Rsun$ for $m_\mathrm{c}\gg 10^4\,\msun$. Note that $\Phi\sim 5.6
\times 10^{5}\bar{f}^{\,-5/3}(4\eta_1)^{-1}\bar{M_1}^{-5/3}$ according to Equation~(\ref{eq:phiGW}).

\section{Discussion}\label{sec:discussion}
\subsection{Summary of results}
We have shown that Advanced LIGO/VIRGO is capable of identifying a third object in the vicinity of a compact object merger by detecting its imprint on the GW waveform. The most prominent perturbation of the third object is due to the time-varying path length to the source (the Doppler phase) as the source orbits around the perturber. Second, the effects of a time-dependent gravitational redshift due to the third object is also significant in many cases. The Shapiro delay may be detectable for intermediate mass (IMBH) or supermassive (SMBH) BH perturbers beyond $10^3\msun$. Dynamical effects of the third object on the orbital elements of the merging binary are less important for circular inspirals unless the pertuber distance is much less than a solar radius. 

The GW Doppler phase may well exceed a radian for a wide range of perturber masses and distances (Figure~\ref{fig:dopplerphase}). For circular NS-NS binaries, a stellar mass compact object companion causes a significant Doppler GW phase shift if it is within a few solar radii ($\sim 10^{11}\,\rm cm$) to the binary and a $10^6\,\msun$ SMBH companion causes a significant Doppler phase if it is within a few AU ($\sim 10^{13}\,\rm cm$). For circular stellar BH-BH binaries the third companion must be a factor $\sim 10$ closer to drive a similar Doppler phase shift (Equation~\ref{eq:phiDsimple}), mainly because the binaries spend a shorter amount of time in the LIGO/VIRGO frequency band (i.e. 16 minutes for circular NS-NS and tens of seconds for circular BH-BH binaries). For these parameters, the effect of the triple companion may be detected in the GW signal as shown by Figure~\ref{fig:SNR} provided that the GW source is within the LIGO/VIRGO horizon (e.g. $S/N\gtrsim 8$ for the unperturbed inspiraling binary). 

\subsection{Event rates and electromagnetic counterparts}\label{sec:eventrates}
The likelihood of discovering such triple systems is currently not well constrained by theoretical models. It is well known that a large fraction of massive stars are in triples (see Section~\ref{sec:introduction}), which may be progenitors of compact object triples detectable by LIGO/VIRGO. However, for known systems, the third object is at a much wider separation than a few solar radii necessary for LIGO/VIRGO detection \citep{2014Natur.505..520R}. The maximum lifetime of close stellar mass compact object triple systems detectable by LIGO/VIRGO is limited by stability arguments and GW emission to within a few Myr (see dashed lines in Figures~\ref{fig:dopplerphase} and \ref{fig:SNR}). These stellar-mass triple systems may form dynamically in dense stellar systems where the encounter rate is high, such as in the cores of globular clusters. Alternatively, a SMBH perturber to a LIGO event may be detected to somewhat larger distances, a few AU (Figure~\ref{fig:SNR}). However most compact object binaries are expected to reside at much larger distances from SMBHs in stellar cusps \citep{2014ApJ...782..101P,2016arXiv160302709S}. The maximum lifetime of these binaries at distances where the SMBH may be detected due to GW emission is a few Myr, similar to stellar-mass perturbers (Figures~\ref{fig:dopplerphase} and \ref{fig:SNR}). Binaries falling to the vicinity of a SMBH may merge due to secular Kozai-Lidov oscillations excited by the SMBH \citep{2012ApJ...757...27A}.

One plausible way to form such tight compact object triples, is in active galactic nuclei (AGN). Compact object binaries may get captured by an accretion disk of a SMBH or form therein. In this case, the SMBH around the binary represents the triple companion. 
The interaction of the binary with the gaseous disk transports the inner binary close to the SMBH, aligns the orbital planes, and drives the inner binary to merge \citep{2016arXiv160203831B,2016arXiv160204226S}. Theoretical estimates of event rates for these mergers is uncertain, estimated to be around a few tens of detections per year for Advanced LIGO/VIRGO. The vicinity of the SMBH may be possibly detected through the Doppler GW phase with LIGO/VIRGO if they migrate to within a thousand gravitational radii of the SMBH. Further in this case, a GW echo may also be possibly detected due to the SMBH  \citep{2013ApJ...763..122K}. Detecting a SMBH triple companion with an aligned orbit with the inner binary may be a smoking gun to infer the presence of an AGN accretion disk in the vicinity of the GW source. Since an inclined outer binary drives nodal precession, an analysis of the GW perturbation driven by the companion may allow one to identify the relative inclination of the inner and outer binaries. The relative inclination may also be measured directly by detecting the GWs of the outer binary with LISA in coincidence (see Section \ref{sec:multiband}).

An attractive property of these GW sources, is that they have electromagnetic counterparts. The accretion disks of AGN are visible to cosmological distances with electromagnetic telescopes and they are much less common than galaxies or globular clusters which allows to cut down on the possible counterpart candidates to the GW event \citep{2006ApJ...637...27K,2008ApJ...684..870K}. 

Furthermore, a massive progenitor star with a short-period compact object binary companion may form a  stellar-mass compact object triple detectable by LIGO/VIRGO. The collapse of a massive star may form a compact object \textit{inner binary}, which merges due to GW emission \citep{2014MNRAS.442.2963K}. In this case the outer object would become the triple companion which leaves its imprints on the GWs of this inner binary. In this case, the collapse of the massive star forming the inner binary might appear as a supernova explosion or a gamma ray burst \citep{2013PhRvL.111o1101R,2016ApJ...819L..21L,2016arXiv160300511W,2016arXiv160404288D}. 

While our estimates were limited to circular-inspiraling inner binaries, we note that triple companions to eccentric inner binaries may be common. The inspiral time of eccentric binaries within the LIGO/VIRGO frequency band may be a factor $\sim100\times$ longer, especially in the highly eccentric, the so-called repeated burst phase \citep{OKL09,2012PhRvD..85l3005K}. For these systems, the triple companion may be at a much larger separation for the binary to execute a significant orbital phase around the triple's center of mass and to cause a significant Doppler GW phase shift for the inner binary signal. Thus, the lifetime of such triples may be much longer, and so the likelihood of detectable triples in LIGO/VIRGO mergers might be expected to be much more common among eccentric LIGO/VIRGO sources. Further, since GW emission tends to decrease the eccentricity as it shrinks the pericenter distance down to merger, the eccentricity may commonly be significant during earlier stages of the inspiral when the GW signal is in the \textit{LISA} frequency band (see Section~\ref{sec:multiband}). Dynamical perturbations of the triple companion may be significant for these sources (Section~\ref{sec:dynamical}). 
Post-Newtonian interaction terms involving all three objects may possibly be detected, which could provide a new test of general relativity \citep{2013ApJ...773..187N,2014PhRvD..89d4043W}. We leave a detailed investigation of eccentric triple GW-sources to future work.

\subsection{Multiband GW detections}\label{sec:multiband}

\begin{figure}
\includegraphics[width=1\columnwidth]{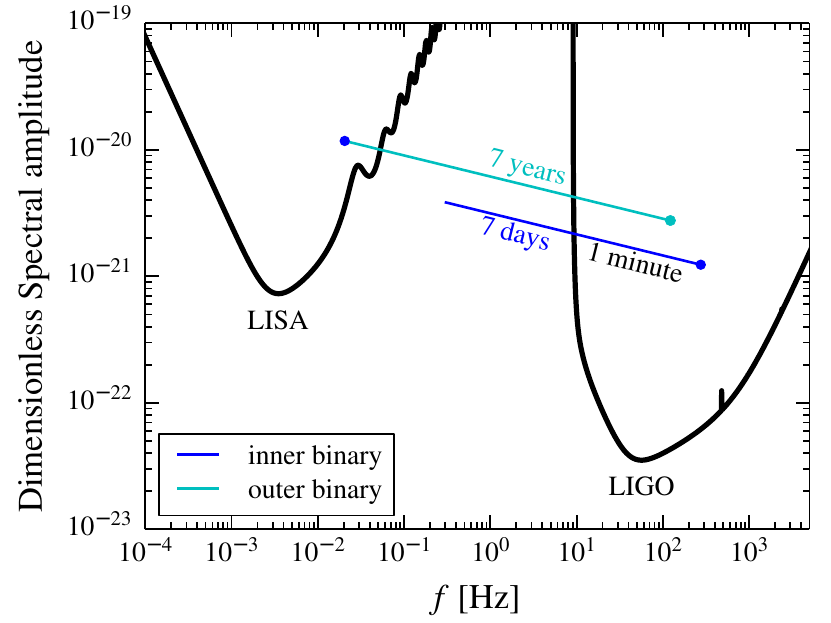}
\caption{\label{fig:multiband}
The dimensionless spectral amplitude due to a triple system comprised of a circular inner binary of two $8\,\msun$ BHs and a $20\,\msun$ companion BH, at a distance of 100\,Mpc. The inner binary is in the LIGO band for 55\,sec, during which the triple companion induces a Doppler shift $\delta \Phi$ of 5.8\,rad. Stability analysis shows that such a circular system must have formed in at most 6.9 days before the inner binary signal was detected with LIGO. The GW frequency of the outer binary is in the \textit{LISA} band during the LIGO detection. Following the inner binary merger, the outer binary GW frequency leaves the \textit{LISA} band and 6.7 years later enters the LIGO band. The filled blue circles represent the ISCO of the inner binary, the cyan circle represents the ISCO of the outer binary (the ringdown is not shown for either binary).}
\end{figure}

Since detecting the presence of a third companion to a merger is primarily limited by the time-duration that the binary spends in the detector's sensitive frequency band, the likelihood of identifying triples may be greatly increased for future GW detectors by decreasing their minimum frequency threshold. Note that the phase shift due to the triple companion scales steeply as $f_{\min}^{-13/3}$ to leading order (Equation~\ref{eq:phiDsimple}). Ultimately, \textit{LISA} will be the best suited to identify stellar mass triples, since here the orbital time of the inner binary may be several years in the detectable frequency band. Note that binaries like  GW150914 could be detected at $S/N \sim 10$ by \textit{LISA} years before merger \citep{2016arXiv160206951S}, and triple companions with orbital periods of years may be possibly discovered for those mergers. 

Is there any other independent way to detect the triple companion in the vicinity of a LIGO/VIRGO source? The orbital frequency of the outer binary, for high SNR LIGO/VIRGO detections, must be comparable to or higher than the inverse merger time of the inner binary in the LIGO/VIRGO frequency band (see thick solid line in Figures~\ref{fig:dopplerphase} and \ref{fig:SNR}). This is $10^{-3}\,{\rm Hz}$ for NS-NS and $0.02\,{\rm Hz}$ for BH-BH binaries, which is well within the sensitive frequency band for \textit{LISA}. Therefore, while the inner binary merges in the LIGO/VIRGO band, the \textit{outer binary} may be coincidentally detected by \textit{LISA}. Further, if the outer binary seperation is sufficiently small, the outer binary itself may merge within a few years following the inner binary merger, which may be detectable with LIGO/VIRGO if the companion mass is less than $10^3\,\msun$. Such spectacular detection sequences may allow for a very accurate parameter estimate determination for these triple systems.

A possible example of such a system is shown in Figure \ref{fig:multiband}, where a $8\,\msun+8\,\msun$ BH-BH circular inner binary is accompanied by a $20\,\msun$ BH on a $0.1\,\Rsun$-separation circular outer orbit, 100\,Mpc from the Earth. If this triple system was circular throughout its prior evolution, then it must have formed within 7 days prior to the inner's merger due to the stability arguments presented in Section~\ref{sec:characteristics}. Thus, the formation of the triple should be captured by \textit{LISA}, as well as the outer binary's inspiral during this phase. This is followed by a detection of the merger event by both \textit{LISA} and LIGO/VIRGO (the total Doppler GW phase shift is 5.8\,rad). Due to the GW recoil kick and a sudden mass loss in the merger process of the inner binary, the outer binary's linear momentum, eccentricity, and inclination suddenly change, leaving an imprint on the outer binary's GW waveform (not shown in the figure) measurable by \textit{LISA}. Following the inner binary coalescence, the inner remnant BH and the outer BH inspirals as an isolated binary, leaving the \textit{LISA} band while continuing to circularize and shrink for 7 years, before showing up in the LIGO/VIRGO band and merging.

\subsection{Search techniques and degeneracies}
Finally, we comment on some practical issues related to data analysis and GW detections. While the number of parameter to describe a binary is generally\footnote{i.e. 9 parameters for circular orbits with nonspinning components, 6 spin parameters, and 2 parameters for eccentric orbits} 17, the number of parameters to fully characterize a hierarchical BH triple system is 27 due to the mass, 6 orbital elements, and 3 spin vector components of the perturber. This may seem to be dauntingly high to carry out a full template-based search for these waveforms. Fortunately, there are several points suggesting that this task may not be impossible. First, the 3 spin components of the third companion do not affect the evolution of the inner binary in any measurable way if the separation is $a_2\gg 10^3 M_2$, since spin effects are higher post-Newtonian order (1.5 PN) \citep{1994PhRvD..49.6274A}. Second, most of the 7 parameters of the triple companion will be degenerate with respect to their effects on the inner binary waveform. For instance, the leading order perturbation, the Doppler phase shift at frequency $f$ (Equation~\ref{eq:phiDsimple}) is set by the line-of-sight acceleration as $\delta \Phi_{\rm D}\propto f^{-10/3}\dot{v}_{1, \rm los}$ approaching merger, where $\dot{v}_{1, \rm los}$ is approximately constant \citep{2011PhRvD..83d4030Y}. The remaining effects are typically much smaller approaching merger. Due to the wide hierarchy in the perturbation effects $\delta \Phi_{\rm D} \gg \delta \Phi_z \gg \delta \Phi_{\rm S}$, and different frequency dependence of these effects, there is room to optimize search algorithms to identify the leading order perturbations of triples as an alternative to a brute force template-based search. 

Importantly, we argue that search algorithms may identify the merging inner binary even if completely neglecting all of the perturbation of the triple. The frequency scaling relations (Eqs.~\ref{eq:Psi} and \ref{eq:phiDsimple}) yield $\delta\Phi_{\rm D}/\Psi = f^{-8/3}$, hence the perturbation is typically negligible at high frequencies approaching the innermost stable circular orbit (ISCO), it accumulates to any substantial level only at much lower frequencies. Indeed, Figure~\ref{fig:spectrum} confirms that the perturbation has a significant $S/N$ per logarithmic frequency interval at frequencies well below the ISCO. The mismatch between an isolated binary waveform and a binary with a triple companion may become significant only below $50\,\rm Hz$ (Figure~\ref{fig:spectrum}). A systematic dephasing at low frequencies could be signs of a third companion. Fortunately, the frequency dependence of the Doppler phase $\delta\Phi_{\rm D}/\Psi = f^{-8/3}$ shows the opposite trends than post-Newtonian corrections which are increasing function of $f$. Thus, a modified mass ratio or spin effects cannot mimic the dephasing associated with a triple companion \citep{2011PhRvD..83d4030Y}. However, the leading order triple companion effects on the waveform may be degenerate with isolated binary waveforms that incorporate possible modifications to the theory of general relativity \citep{2016arXiv160203841T,2016arXiv160308955Y}.

\acknowledgments{
This work was supported in part by the European Research Council under the European Union's Horizon 2020 Programme, ERC-2014-STG grant GalNUC 638435 and by NSF grant AST-1312034. The calculations were carried out on the NIIF HPC cluster at the University of Debrecen, Hungary.}

\bibliographystyle{yahapj}
\bibliography{main}

\end{document}